\documentstyle[12pt,bezier]{article}
\begin{document}
\def \inbar{\vrule height1.5ex width.4pt depth0pt}
\def \xC{\relax\hbox{\kern.25em$\inbar\kern-.3em{\rm C}$}}
\def \xR{\relax{\rm I\kern-.18em R}}
\newcommand{\R}{\xR}
\newcommand{\C}{\xC}
\newcommand{\xZ}{Z \hspace{-.08in}Z}
\newcommand{\Z}{Z \hspace{-.08in}Z}
\newcommand{\xbe}{\begin{equation}}
\newcommand{\be}{\begin{equation}}
\newcommand{\xee}{\end{equation}}
\newcommand{\ee}{\end{equation}}
\newcommand{\xbea}{\begin{eqnarray}}
\newcommand{\bea}{\begin{eqnarray}}
\newcommand{\xeea}{\end{eqnarray}}
\newcommand{\eea}{\end{eqnarray}}
\newcommand{\xnn}{\nonumber}
\newcommand{\nn}{\nonumber}
\newcommand{\xkt}{\rangle}
\newcommand{\kt}{\rangle}
\newcommand{\xbr}{\langle}
\newcommand{\br}{\langle}
\newcommand{\xcun}{\mbox{\footnotesize${\cal N}$}}
\newcommand{\cun}{\mbox{\footnotesize${\cal N}$}}
\newcommand{\cum}{\mbox{\footnotesize${\cal M}$}}
\newcommand{\Q}{{\cal Q}}
\title{Topological Symmetries}
\author{A.~Mostafazadeh$ ^{a}$ and 
K.~Aghababaei Samani$ ^{b}$\\ \\
$ ^{a}$~Department of Mathematics, Ko\c{c} University,\\
Istinye 80860, Istanbul, TURKEY\thanks{E-mail address: 
amostafazadeh@ku.edu.tr}\\ 
$ ^{b}$~Institute for Advanced Studies in Basic Sciences,\\
45195-159 Gava Zang, Zanjan, IRAN}
\date{ }
\maketitle

\begin{abstract}
We introduce the notion of a topological symmetry as a 
quantum mechanical symmetry involving a certain topological 
invariant. We obtain the underlying algebraic structure of 
the $\Z_2$-graded uniform topological symmetries of type 
$(1,1)$ and $(2,1)$. This leads to a novel derivation of the 
algebras of  supersymmetry and $p=2$ parasupersummetry. 
\end{abstract}


\baselineskip=24pt

\section{Introduction} 
Since Witten's pioneering work on supersymmetric quantum mechanics 
\cite{witten-82}, there has been a growing interest in supersymmetry 
and its applications \cite{susy,review,junker}. The interest in supersymmetry
has also motivated the development of various geralizations of supersymmetry. 
Most notable of these are parasupersymmetries
\cite{ru-sp,be-de,khare}, the $q$-deformed supersymmetries \cite{def}
and the fractional supersymmetries \cite{frac}. 

One of the most intriguing aspects of supersymmetric quantum mechanics 
is its relationship with the Atiyah-Singer index theorem \cite{index-thm}. 
This has already been noticed by Witten \cite{witten-82} in early 1980's 
and subsequently led to new proofs of this theorem 
\cite{alvarez-gaume,windey,jmp94}. Supersymmetric proofs of the index 
theorem together with Witten's supersymmetric derivation of the Morse 
inequalities \cite{morse} and its impact on Floer's theory 
\cite{floer,nash} are among the greatest mathematical achievments of 
supersymmetric quantum mechanics. 

The effectiveness of supersymmetry in providing insight into some of 
the most fundamental results of differential geometry and topology 
suggests the study of the topological content of various generalizations 
of supersymmetry. The only known generalization of supersymmetry which 
exhibits similar topological properties is a certain type of $p=2$ 
parasupersymmetries. The topological properties of $p=2$ parasupersymmetry 
(PSUSY) has been extensively studied in Refs.~\cite{ijmpa-96a,ijmpa-97}.

The purpose of the present article is to introduce a generalization of 
supersymmetry (SUSY) which shares its topological properties. We shall
term such a symmetry a {\em topological quantum mechanical symmetry} 
or simply {\em topological symmetry} (TS).  

The first step towards such a generalization is to recall the basic properties
of the $N=1$ supersymmetric quantum mechanics.\footnote{We use the terminology 
in which $N$ is half of the number of Hermitian generators. What we call 
$N=1$ SUSY sometimes is referred to as the ${\cal N}=2$ SUSY where ${\cal N}$ 
is the number of Hermitian SUSY generators.} 

$N=1$ supersymmetric quantum mechanics is specified by a $\Z_2$-graded 
Hilbert space ${\cal H}={\cal H}_+\oplus{\cal H}_-$ and the superalgebra
	\bea
	\left[ H,\Q\right]&=&0\;,
	\label{2.1}\\
	H&=&\frac{1}{2}\;\{\Q,\Q^\dagger\}\;,
	\label{2.2}\\
	\Q^2&=&0\;,
	\label{2.3}
	\eea
where $H$ and $\Q$ are the Hamiltonian and the generator of the 
supersymmetry, respectively. The $\Z_2$-grading of the Hilbert space is
implemented using a  chirality or parity operator $\tau:{\cal H}\to{\cal H}$ satisfying
	\bea
	\tau^2&=&1\;,~~~\tau^\dagger=\tau\;,
	\label{1.4}\\
	\left[\tau,H\right]&=&0\;,
	\label{1.5}\\
	\{\tau,\Q\}&=&0\;.
	\label{1.6}
	\eea
The subspaces ${\cal H}_+$ and ${\cal H}_-$ are identified with
the eigenspaces of $\tau$, 
	\be
	{\cal H}_\pm:=\{ |\psi\kt\in{\cal H}~|~\tau|\psi\kt=
	\pm|\psi\kt\}\;.
	\label{2.4}
	\ee
The elements of ${\cal H}_\pm$ are said to have definite 
parity $\pm$. An operator $O$ acting on ${\cal H}$ is said to have 
definite parity $+$ or $-$, if it commutes or anticommutes with $\tau$, 
respectively. 

It is well-known that using Eqs.~(\ref{2.1}) -- (\ref{2.3}) one can 
obtain the degeneracy structure of a general $N=1$ supersymmetric 
system \cite{witten-82,susy,ijmpa-96a}. These systems have a 
nonnegative spectrum and the eigenspaces corresponding to positive 
eigenvalues are spanned by pairs of eigenvectors of opposite parity. 
This particular degeneracy structure of supersymmetric systems is 
sufficient to show the topological invariance of the Witten index: 
	\bea
	&&{\rm index}_{\rm W}:=n_0^+-n_0^-\;,
	\label{2.5}\\
	&&n_0^\pm:=
	{\rm number~of~zero~energy~states~with~parity~} \pm\;.
	\label{2.6}
	\eea
The degeneracy structure is obtained from the algebraic structure. 
Therefore, the situation may be described by the following diagram.
	\be
	{\rm algebraic~structure}~\to~{\rm degeneracy~structure}~
	\to~{\rm topological~invariants}
	\label{2.8}
	\ee
The same analysis is valid for the case of  the $p=2$ PSUSY studied in
Refs.~\cite{ijmpa-96a,ijmpa-97}.

The idea pursued in this article is to reverse the first arrow in 
(\ref{2.8}). More specifically, we wish to 
\begin{itemize}
\item
find and postulate the type of degeneracy structures which lead to 
topological invariants such as the Witten index, and
\item
obtain the algebraic structure of symmetries which support this 
type of degeneracy structures.
\end{itemize}

\section{Topological Symmetries and Their Invariants}

\begin{itemize}
\item[~] {\bf Definition~1:} Let $m_+$ and $m_-$ be two positive 
integers. Then a quantum mechanical symmetry is called a 
{\em $\Z_2$-graded topological symmetry} (TS) of type 
$(m_+,m_-)$, if the following conditions are fulfilled.
\begin{itemize}
\item[1.]
The Hilbert space is $\Z_2$-graded. The $\Z_2$-grading is achieved 
via a parity operator $\tau$ satisfying Eqs.~(\ref{1.4}), i.e., 
${\cal H}={\cal H}_+\oplus {\cal H}_-$ where ${\cal H}_\pm$ are 
given by Eq.~(\ref{2.4}).
\item[2.]
The energy spectrum of the system is nonnegative.
\item[3.] There is an energy eigenbasis consisting of definite parity
state vectors, i.e., Eq.~(\ref{1.5}) holds. 
\item[4.]
For every positive energy eigenvalue E, there exists a positive
integer $\lambda_E$ such that $E$ is $m_E:=\lambda_E(m_-+m_+)$ fold 
degenerate. Furthermore, the corresponding eigenspace is spanned by
$\lambda_Em_-$ negative parity eigenvectors and $\lambda_Em_+$ positive
parity eigenvectors.
\end{itemize}
\end{itemize}
Clearly, SUSY is an example of TS of type $(1,1)$. Therefore, TSs are 
generalizations of the SUSY.

\begin{itemize}
\item[~] {\bf Definition~2:} A $\Z_2$-graded topological symmetry is said
to be {\em uniform}, if for all $E>0$, $\lambda_E=1$.
\end{itemize}

Given the above definition of a $\Z_2$-graded uniform topological 
symmetry (UTS), we can easily prove the following theorem.
\begin{itemize}
\item[~] {\bf Theorem:} Consider a  $\Z_2$-graded UTS of type $(m_-,m_+)$
and let $n^\pm_0$ denote the number of zero energy eigenstates 
of the system with parity $\pm$. Then the quantity
	\be
	\Delta_{(m_+,m_-)}:=m_-n_0^+-m_+n_0^-
	\label{3.1}
	\ee
is a topological invariant, i.e., it is invariant under the
continuous changes of the quantum system\footnote{These include
continuous changes of the Hamiltonian and the boundary 
conditions.} that do not destroy the UTS.
\item[~] {\bf Proof:} Suppose that under a continuous change of the system
a zero energy state vector with positive parity is elevated to a positive
energy level. This positive energy level must have $m_+$ positive parity
states and $m_-$ negative parity states. Hence, the initial zero energy state
must be accompanied by $(m_+-1)$ positive parity zero energy eigenstates
and $m_-$ negative parity zero energy eigenstates. This implies that after
the change $\Delta_{(m_+,m_-)}$ is given by
	\[\Delta_{(m_+,m_-)}^{\rm after}=
	m_-(n_0^+-m_+)-m_+(n_0^--m_-)=m_-n_0^+-m_+n_0^-=
	\Delta_{(m_+,m_-)}^{\rm before}\;.\]
Every possible change of the zero energy states is a combination 
of this particular change and its converse. Therefore, in general 
$\Delta_{(m_+,m_-)}$ remains invariant.~$\Box$
\end{itemize}
The same proof is valid for the case of nonuniform $\Z_2$-graded TSs. 
Therefore, $\Delta_{(m_+,m_-)}$ is a topological invariant for any
$\Z_2$-graded TS.

We shall next provide the basic framework for addressing the 
characterization problem for TSs. In this paper we shall only
consider the uniform TSs. But our method applies to nonuniform
TSs as well.

We shall demand TSs to have symmetry generators $\Q_a$ (with 
$a=1,2,\cdots,N$) which have negative parity. In 
particular, we shall only consider the $N=1$ UTSs where the 
label $a=1$ can be dropped. In this case, Eq.~(\ref{1.6}) is valid.

Next we introduce the Hermitian symmetry generators,
	\be
	Q_1:=\frac{1}{\sqrt{2}}(\Q+\Q^\dagger)\;,~~~
	{\rm and}~~~Q_2=\frac{-i}{\sqrt{2}}(\Q-\Q^\dagger)\;.
	\label{3.2}
	\ee
In view of Eqs.~(\ref{2.1}),  (\ref{1.4}), (\ref{1.6}) and (\ref{3.2}), we have
	\bea
	\left[H,Q_i\right]&=&0\;,
	\label{3.3}\\
	\left\{\tau,Q_i\right\}&=&0\;,
	\label{3.5}
	\eea
where $i\in\{1,2\}$.

We can use Eq.~(\ref{1.5}) to construct an orthonormal basis of 
the Hilbert space in which $H$ and $\tau$ are diagonal. Our 
strategy will be to use the information on the degeneracy 
structure of the energy eigenspaces and Eqs.~(\ref{1.4}), 
(\ref{3.3}), and (\ref{3.5}) to obtain matrix representations of 
$Q_1$ and $Q_2$ in the energy eigenspaces ${\cal H}_E$ with 
eigenvalue $E>0$. We shall denote the representation of an operator 
$O$ in the eigenspace ${\cal H}_E$ by $O^E$. Clearly for the 
Hamiltonian $H$, we have $H^E=E I_m$, where $I_m$ is the $m\times m$ 
identity matrix and $m:=m_++m_-$. Finally, we shall try to use these 
representations to obtain the most general algebraic relations 
satisfied by $\Q$ and $H$.

\section{Uniform Topological Symmetries of Type $(1,1)$}

For the $\Z_2$-graded TS of type $(1,1)$, the positive energy levels are
doubly degenerate ($m=2$). In a basis that diagonalizes $H$ and $\tau$, 
we have (up to a permutation of the basis vectors)
	\be
	\tau^E=\left(\begin{array}{cc}
	1&0\\
	0&-1\end{array}\right)\;,
 	\label{n4.1}
	\ee
where $E>0$ and we have used Eqs.~(\ref{1.4}). Next let us note that
$Q_i^E$ are $2\times 2$ Hermitian matrices satisfying (\ref{3.5}).
These conditions are sufficient to conclude that
	\be
	Q_1^E=\left(\begin{array}{cc}
	0&\mu^*\\
	\mu&0\end{array}\right)~~~{\rm and}~~~
	Q_2^E=\left(\begin{array}{cc}
	0&\nu^*\\
	\nu&0\end{array}\right)\;,
 	\label{n4.2}
	\ee
where $\mu$ and $\nu$ are complex numbers. 

Using the matrix representations (\ref{n4.1}) and (\ref{n4.2}), we can
easily compute
	\bea
	\Q^E&:=&\frac{1}{\sqrt{2}}\,(Q_1^E+iQ_2^E)=\frac{1}{\sqrt{2}}
	\:\left(\begin{array}{cc}
	0&\mu^*+i\nu^*\\
	\mu+i\nu&0\end{array}\right)\;,
	\label{n4.3}\\
	(Q_1^E)^2&=&|\mu|^2 I_2\;,
	\label{n4.4}\\
	(Q_2^E)^2&=&|\nu|^2 I_2\;,
	\label{n4.5}\\
	(\Q^E)^2&=&\frac{1}{2}\:\left[ |\mu|^2-|\nu|^2+
	i(\mu\nu^*+\mu^*\nu)\right] I_2=-\det(\Q^E)I_2\;,
	\label{n4.6}
	\eea
where `$\det$' stands for `determinant'. Next we introduce
the Hermitian operators $M$, $K_1$ and $K_2$ which commute
with $H$ and have the following representations in ${\cal H}_E$
with $E>0$.
	\be
	M^E=|\mu|^2~I_2\,,~~~K_1^E=(|\mu|^2-|\nu|^2)I_2\,,~~~
	K_2^E=(\mu\nu^*+\mu^*\nu)I_2\;.
	\label{n4.9}
	\ee
In view of Eqs.~(\ref{n4.4}) -- (\ref{n4.9}), $M^E$, $K^E$, 
and $K_i^E$ commute with $Q_i^E,\Q^E$ and $\tau^E$. Generalizaing
these equations to operator identities, we find
	\bea
	&&[M,Q_i]=[K_j,Q_i]=[M,K_j]=0\;,
	\label{a0}\\
	&&Q_1^2=M\;,
	\label{a1}\\
	&&Q_2^2=M-K_1\;,
	\label{a2}\\
	&&\{Q_1,Q_2\}=K_2\;.
	\label{a3}
	\eea
We can also express Eqs.~(\ref{a0}) -- (\ref{a3}) in terms of $\Q$.
This yields
	\bea
	&&[M,\Q]=[K,\Q]=[M,K]=0\;,
	\label{a4}\\
	&&\frac{1}{2}\,\{\Q,\Q^\dagger\}=M-\frac{1}{2}(K+K^\dagger)\;,
	\label{a5}\\
	&&\Q^2=K\;,
	\label{a6}
	\eea
where $K:=(K_1+iK_2)/2$.
 
Next let us note that under a linear transformations of the form:
	\be
	\begin{array}{ccc}
	Q_1&\to&\tilde Q_1:=a\: Q_1+b\: Q_2\;,\\
	Q_2&\to&\tilde Q_2:=c\: Q_1+d\: Q_2\;,
	\end{array}
	\label{n4.10}
	\ee
the algebra (\ref{a0}) -- (\ref{a3}) is left form-invariant.\footnote{Here
the coefficients $a,b,c$ and $d$ are assumed to be Hermitian operators commuting
with $H$, $Q_i$, $K_i$ and $M$.} More specifically, the transformed generators 
$\tilde Q_i$ satisfy the same algebra provided that the operators $M$ and
$K_i$ are transformed according to
	\bea
	M&\to&\tilde M:=(a^2+b^2)M-b^2K_1+abK_2\;,
	\label{t1}\\
	K_1&\to&\tilde K_1:=(a^2+b^2-c^2-d^2)M+(d^2-b^2)K_1+
	(ab-cd)K_2\;,
	\label{t2}\\
	K_2&\to&\tilde K_2:=2(ac+bd)M-2bd K_1+(ad+bc)K_2\;.
	\label{t3}
	\eea
This observation may be used to find a new set of negative parity 
symmetry generators $\tilde Q_i$ which would satisfy the algebra
(\ref{a0}) -- (\ref{a3}) with $K_i$ set to zero. The most general linear
transformations (\ref{n4.10}) for which $\tilde K_1=\tilde K_2=0$ are the 
ones satisfying (either of)
	\be
	\frac{a+ic}{b+id}=-\frac{K_2}{2M}\pm i\sqrt{1-\frac{K_1}{M}-
	\frac{K_2^2}{4M^2}}\;.
	\label{n4.12}
	\ee
Using the energy eigenbasis in which Eqs.~(\ref{n4.9}) hold, we can
show that indeed the terms inside the square root in (\ref{n4.12}) add up to
a positive operator.\footnote{Strictly speaking most of the conclusions 
drawn in this article are based on the information obtained from the
restriction of the relevant operators to ${\cal H}-{\cal H}_0$. We 
will however suppose that the same results are generally valid in
${\cal H}$. This is consistent with the fact that we have not given
the representations of $M$ and $K_i$ in ${\cal H}_0$.} This is a
necessary condition for the corresponding linear transformation to be
invertible.

The above analysis shows that without loss of generality we can set 
$K=0$ in Eqs.~(\ref{a4}) -- (\ref{a6}). This yields
	\bea
	&&[M,\Q]=0\;,
	\label{a7}\\
	&&\frac{1}{2}\{\Q,\Q^\dagger\}=M\;,
	\label{a8}\\
	&&\Q^2=0\;.
	\label{a9}
	\eea
Since $M$ has the same degeneracy structure as the Hamiltonian (at least
in ${\cal H}-{\cal H}_0$), we can write $H$ as a function of $M$.
In particular, we can identify $H$ with $M$, in which case the algebra
(\ref{a7}) -- (\ref{a9}) reduces to the superalgebra (\ref{2.1}) --
(\ref{2.3}). Therefore, the algebra of the $\Z_2$-graded UTS of type
$(1,1)$ is the same as the algebra of SUSY. The above analysis may be
viewed as a derivation of the superalgebra (\ref{2.1}) -- (\ref{2.3}) 
from a set of basic principles, i.e., the definition of the $\Z_2$-graded
UTS of type $(1,1)$.  

\section{Uniform Topological Symmetry of Type $(2,1)$}

For the $\Z_2$-graded UTS of type $(2,1)$, the positive energy levels are
triply degenerate ($m=3$). We will again work in a basis in which $H$ 
and $\tau$ are diagonal.  Restricting to an eigenspace ${\cal H}_E$ 
of $H$ with $E>0$ and enforcing Eqs.~(\ref{1.4}) and (\ref{3.5}), 
we can easily show that (up to permutations of the basis vectors of 
${\cal H}_E$)
	\bea
	&&\tau^E=\left(\begin{array}{ccc}
	1&0&0\\
	0&1&0\\
	0&0&-1
	\end{array}\right)\,,~~~~
	Q_1^E=\left(\begin{array}{ccc}
	0&0&\mu_1^*\\
	0&0&\mu_2^*\\
	\mu_1&\mu_2&0
	\end{array}\right)\,,
	\label{n6.1}\\
	&&Q_2^E=\left(\begin{array}{ccc}
	0&0&\nu_1^*\\
	0&0&\nu_2^*\\
	\nu_1&\nu_2&0
	\end{array}\right)\,,~~~~
	\Q^E=\frac{1}{\sqrt{2}}\left(\begin{array}{ccc}
	0&0&\mu_1^*+i\nu_1^*\\
	0&0&\mu_2^*+i\nu_2^*\\
	\mu_1+i\nu_1&\mu_2+i\nu_2&0
	\end{array}\right)\,.
	\label{n6.2}
	\eea

Now in order to obtain the most general algebraic identities satisfied by
these matrices we appeal to the Cayley-Hamilton theorem 
\cite{cayley-hamilton}. This theorem states that any $m\times m$ matrix $A$ 
satisfies its characteristic equation, $P_A(x)=0$, where $P_A(x)$ is the 
characteristic polynomial for $A$. It is not difficult to show that the
characteristic polynomial for a $3\times 3$ matrix $A$ of the form
	\[A=\left(\begin{array}{ccc}
	0&0&\alpha\\
	0&0&\beta\\
	\gamma&\delta&0
	\end{array}\right)\]
is given by $P_A(x)=x^3-(\alpha\gamma+\beta\delta)x$. Using this equation
and the identity $P_A(A)=0$ for $Q^E_1$, $Q^E_2$ and $\Q^E$, we find
	\bea
	(Q^E_1)^3&=& \sum_{j=1}^2|\mu_j|^2~Q_1^E\;,
	\label{6.14}\\
	(Q^E_2)^3&=&\sum_{j=1}^2|\nu_j|^2~Q^E_2\;,
	\label{6.15}\\
	(\Q^E)^3&=&\frac{1}{2}\,\sum_{j=1}^2[|\mu_j|^2-|\nu_j|^2+
	i(\mu_j\nu_j^*+\mu_j^*\nu_j)]~\Q^E\;.
	\label{6.16}
	\eea

Next we introduce the Hermitian operators $M$, $K_1$, and $K_2$
which commute with $H$ and have the following matrix 
representations in ${\cal H}_E$ with $E>0$.
	\bea
	M^E&=&\sum_{j=1}^2|\mu_j|^2~I_3\;,
	\label{n6.17}\\
	K_1^E&=&\sum_{j=1}^2(|\mu_j|^2-|\nu_j|^2)~I_3\;,
	\label{n6.18}\\
	K_2^E&=&\sum_{j=1}^2(\mu_j\nu_j^*+\mu_j^*\nu_j)~I_3\;.
	\label{n6.19}
	\eea
Now setting $K:=(K_1+iK_2)/2$ and making use of Eqs.~(\ref{n6.17})
-- (\ref{n6.19}), we can generalize Eqs.~(\ref{6.14}) -- (\ref{6.16})
to the operator relations
	\bea
	Q_1^3&=&MQ_1\;,
	\label{6.23}\\
	Q_2^3&=&(M-K_1)Q_2\;,
	\label{6.24}\\
	\Q^3&=&K\,\Q\;.
	\label{6.25}
	\eea
Eqs.~(\ref{n6.17}) -- (\ref{n6.19}) also suggest that the operators
$M$ and $K_i$ commute among themselves and with $Q_i$. Furthermore, we can rewrite
Eq.~(\ref{6.25}) in terms of $Q_1$ and $Q_2$. Simplifying the resulting
equation using Eqs.~(\ref{6.23}) and (\ref{6.24}) we obtain the 
following general algebra
	\bea
	&&[M,Q_i]=[M,K_i]=[K_i,Q_j]=0\;,
	\label{b0}\\
	&&Q_1^3=MQ_1\;,
	\label{b1}\\
	&&Q_2^3=(M-K_1)Q_2\;,
	\label{b2}\\
	&&Q_2Q_1Q_2+\{Q_1,Q_2^2\}=(M-K_1)Q_1+K_2Q_2\;,
	\label{b3}\\
	&&Q_1Q_2Q_1+\{Q_2,Q_1^2\}=MQ_2+K_2Q_1\;.
	\label{b4}
	\eea

We can show by direct computation that this algebra remains 
form-invariant under linear transformations (\ref{n4.10}) of $Q_i$.
More remarkable is the fact that under such a transformation the
operators $M$ and $K_i$ transform according to the same relations as
in the case of UTS of type $(1,1)$, namely Eqs.~(\ref{t1}) -- (\ref{t3}).
Therefore, we can always transform to a new set of symmetry
generators for which $K_i=0$. Setting $K_i=0$ in Eqs.~(\ref{b0}) --
(\ref{b4}) and rewriting these equations in terms of $\Q$, we find
	\bea
	&&[M,\Q]=0\;,
	\label{6.50}\\
	&&\{\Q^2,\Q^\dagger\}+\Q\Q^\dagger\Q=2M\Q\;,
	\label{6.51}\\
	&&\Q^3=0\;.
	\label{6.52}
	\eea

The operator $M$ has the same degeneracy structure as the Hamiltonian. 
Therefore, we can identify $H$ with a function of $M$. In particular, 
we can set $H=M/2$. In this case the algebra (\ref{6.50}) -- (\ref{6.52}) 
becomes identical to the algebra of $p=2$ parasupersymmetry 
\cite{ru-sp}. In other words, the algebra of the $\Z_2$-graded UTS of 
type $(2,1)$ is precisely the algebra of the $p=2$ parasupersymmetry 
(of Rubakov and Spiridonov \cite{ru-sp}). Again, the above analysis may 
be viewed as a derivation of the algebra of $p=2$ parasupersymmetry from 
a set of basic principles, i.e., the definition of the $\Z_2$-graded UTS 
of type $(2,1)$.
 
As shown in Ref.~\cite{ijmpa-96a}, the algebra (\ref{6.50}) -- 
(\ref{6.52}) does not imply the degeneracy structure of type $(2,1)$ 
UTS. Therefore, the type $(2,1)$ $\Z_2$-graded UTSs belong to a 
special class of symmetries whose generator $\Q$ satisfies 
Eqs.~(\ref{6.50}) -- (\ref{6.52}). A method for constructing
the corresponding moduli spaces is given in Ref.~\cite{ijmpa-97}.

\section{Conclusion}
In this article we have introduced a general notion of a topological 
symmetry.  We have provided a simple framework for the study
of the $\Z_2$-graded topological symmetries.  We showed that 
the algebras of the $\Z_2$-graded UTS of order $(1,1)$ and $(2,1)$ 
are essentially the algebras of supersymmetric quantum mechanics and 
$p=2$ parasupersymmetric quantum mechanics, respectively.

By construction, topological symmetries involve a class of 
integer-valued topological invariants $\Delta_{(m_+,m_-)}$. These are 
the analogues of the Witten index of supersymmetry and the 
parasupersymmetric topological invariant \cite{ijmpa-97}. The physical 
interpretation of these invariants is that they are a measure of the 
existence of zero-energy ground states. For the known cases, the latter
is an indication of the exactness of symmetry. The mathematical 
interpretation of $\Delta_{(m_+,m_-)}$ is not quite clear. For the 
known cases they are related to the analytic indices of Fredholm
operators \cite{ijmpa-97}. The general case requires a detailed
study of the algebraic structure of general topological symmetries. 
The algebras of $\Z_2$-graded TS of arbitrary type $(m_+,m_-)$ are 
currently under investigation.

Finally, one can easily generalize the definition of the $\Z_2$-graded
topological symmetry of type $(m_+,m_-)$ to a $\Z_n$-graded topological
symmetry of type $(m_1,m_2,\cdots,m_n)$. Such a system will have states
with definite `color' taking values in $\{1,2,\cdots,n\}$. The spectrum 
will be nonnegative. The positive energy eigenvalues $E$ will be 
$m_E=\lambda_E\sum_{\ell=1}^nm_\ell$ fold degenerate. The energy eigenspaces with 
positive eigenvalue will all have $\lambda_Em_1$ states of color $1$, 
$\lambda_Em_2$ states of color $2$, $\cdots$, and $\lambda_E m_n$ states
of color $n$. One may try to define topological invariants for these more 
general topological symmetries. For example for a $\Z_3$-graded TS
of type $(1,1,1)$, one can introduce the invariant $\Delta_{(1,1,1)}=
(n_0^{(1)}-n_0^{(2)})^2+(n_0^{(2)}-n_0^{(3)})^2+(n_0^{(3)}-n_0^{(1)})^2$, 
where $n_0^{(\ell)}$ is the number of zero energy states of color $\ell$.
The basic properties of $\Z_n$-graded topological symmetries will be
explored in \cite{p34b}.

\end{document}